# Resonant Readout of a Persistent Current Qubit

Janice C. Lee, William D. Oliver, Terry P. Orlando, and Karl K. Berggren

*Abstract*— We have implemented a resonant circuit that uses a SQUID as a flux-sensitive Josephson inductor for qubit readout. In contrast to the conventional switching current measurement that generates undesired quasi-particles when the SQUID switches to the voltage state, our approach keeps the readout SQUID biased along the supercurrent branch during the measurement. By incorporating the SQUID inductor in a high-Q resonant circuit, we can distinguish the two flux states of a niobium persistent-current (PC) qubit by observing a shift in the resonant frequency of both the magnitude and the phase spectra. The readout circuit was also characterized in the nonlinear regime to investigate its potential use as a nonlinear amplifier.

## I. Introduction

Q UANTUM computation with superconducting flux qubits commonly relies on DC SQUID magnetometers for qubit readout. While the coherent control of quantum dynamics has been demonstrated in several flux-qubit systems [1]-[3], a major challenge has been to reduce the level of decoherence of the qubit state caused by the readout circuit. In the case of persistent-current qubits, the conventional SQUID switching-current readout has several drawbacks. To distinguish the flux state of the qubit, one typically ramps the current bias through the SQUID and records the presence or absence of a switching event. Decoherence due to this readout process occurs mainly in three ways: (1) non-zero SQUID currents result in SQUID-potential asymmetries, which remove the first-order current-noise isolation and are thereby associated with higher levels of decoherence (e.g. through the mutual coupling of broadband noise from the SQUID bias lines to the qubit loop) [4]; (2) the generation of quasi-particles from the SQUID switching events limits the repetition rate of the readout; and (3) the undesired excitation of the qubit into higher energy states during the ramping or switching of the SQUID.

To address in part these sources of readout-induced decoherence, we have experimentally implemented a resonant readout technique that only requires the readout SQUID to be biased at low currents along the supercurrent branch. The low current bias tends to maintain the first-order noise isolation, helping to minimize the level of decoherence of the qubit. Since the SQUID does not switch to the voltage state, the number of quasi-particles is also drastically reduced. In addition, the resonant readout approach utilizes a narrow-band filter that shields the qubit from broadband noise. Similar resonant readout approaches have been demonstrated in Refs. [5] and [6], and the implementation has been extended to the nonlinear regime in Ref. [7].

The principle of the resonant readout technique for the persistent-current qubit is to determine the flux state of the qubit by measuring the Josephson inductance of the readout SQUID. As illustrated in Fig. 1, the SQUID Josephson inductance is nonlinear: it is a nonlinear and periodic function of the external magnetic flux (used here to detect the qubit state), as well as a nonlinear function of the bias current (used here to control the degree of nonlinearity in the resonant circuit). To measure the Josephson inductance with high sensitivity, we incorporate the SQUID inductor in a high-Q resonant circuit. The change in Josephson inductance corresponding to a transition between qubit states is detected as a shift in the resonant frequency (Fig. 2).

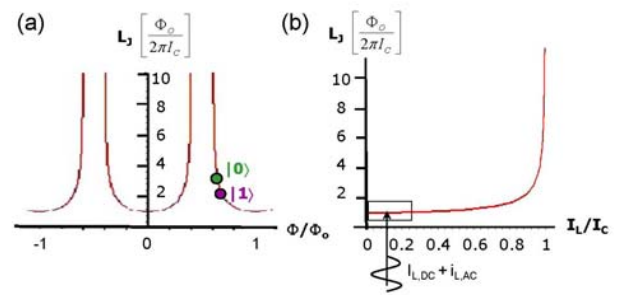

Manuscript received October 5, 2004. This work was supported in part by the AFOSR grant F49620-01-1-0457 under the DoD University Research Initiative on Nanotechnology (DURINT) Program, by an NSF graduate fellowship, and MIT Lincoln Laboratory by Department of the Air Force under Air Force Contract F19628-00-C-0002. Opinions, interpretations, conclusions, and recommendations are those of the author(s) and are not necessarily endorsed by the United States government.

J. C. Lee and T. P. Orlando are with the Department of Electrical Engineering and Computer Science, Massachusetts Institute of Technology, Cambridge, MA 02139 USA (phone: 617-253-0393; fax: 617-258-6640; e-mail: janlee@mit.edu).
W. D. Oliver is with MIT Lincoln Laboratory, Lexington, MA 02420 USA (email: oliver@ll.mit.edu).
K. K. Berggren was with MIT Lincoln Laboratory, Lexington, MA 02420 USA. He is now with the Department of Electrical and Computer Engineering, Massachusetts Institute of Technology, Cambridge, MA 02139 USA (email: berggren@mit.edu).

Fig. 1. a) The SQUID Josephson inductance is a periodic function of the external magnetic flux. The figure shows the qubit-mediated Josephson inductance for the two flux states ($|0\rangle$ and $|1\rangle$) at a flux bias near 0.7 $\Phi_o$. b) Dependence of the Josephson inductance on the bias current. A small current bias corresponds to a linear inductance, i.e., independent of current (boxed region). Increasing current bias causes the SQUID inductor to behave nonlinearly, which has observable effects on the resonant spectrum.

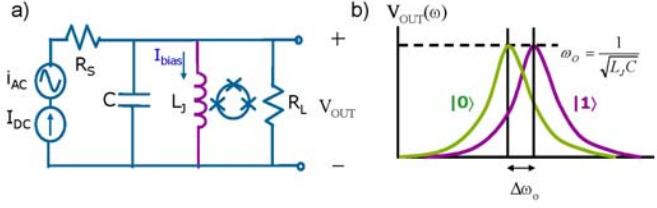

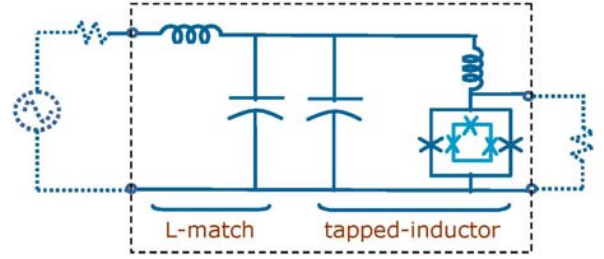

Fig. 2. a) A generic parallel RLC circuit, in which the inductor is realized by a SQUID, with its Josephson inductance being mediated by the qubit. b) The transition between qubit states is detected as a shift in the resonant frequency.

## II. RESONANT CIRCUIT DESIGN AND FABRICATION

### A. On-Chip Resonant Circuit Design

The circuit design of the resonant readout circuit is shown in Fig. 3. The components within the dotted box were implemented on-chip. The qubit is inductively coupled to the readout SQUID inductor, which in turn is incorporated in a parallel resonant circuit. To achieve a high quality factor, a tapped-inductor impedance transformation was employed to step-up the output impedance seen by the oscillator at its resonant frequency. The SQUID inductor is part of this tapped-inductor configuration. An L-match network was also used on the input side to match the input impedance to the transformed output impedance. The transformations were designed to yield a resonant frequency $f_0 \sim 500$ MHz and a quality factor $Q \sim 150$ [8].

### B. Device Fabrication

The device was fabricated with a planarized Nb trilayer process at MIT Lincoln Laboratory [9]. The PC qubit is a superconducting loop interrupted by three Josephson junctions. Two of the junctions have the same size with nominal dimensions of $1.0 \times 1.0$ $\mu m^2$. The third junction is smaller with nominal dimensions of $0.9 \times 0.9$ $\mu m^2$. The junctions of the readout SQUID are $1.5 \times 1.5$ $\mu m^2$. Based on the results of the process tests, the current density was approximately 120 A/cm$^2$. The dimensions of the qubit loop and the SQUID loop are $18.0 \times 18.0$ $\mu m^2$ and $20.8 \times 20.8$ $\mu m^2$ respectively, with an estimated mutual inductance of 30 pH. The inductors were square spirals with a linewidth and spacing of 1μm. The capacitors were made out of Nb electrodes, with the dielectric layers being 50 nm of NbO$_x$ and 200nm of SiO$_2$. A device micrograph is shown in Fig. 4.

Fig. 3. Circuit schematic of the resonant readout circuit. Tapped-inductor and L-match impedance transformations were employed to achieve a high quality factor. The component values are $L_1$ = 69 nH, $C_1$ = 1.4pF, $C_2$ = 100 pF, $L_2$ = 0.78 nH. The junctions of the SQUID are each shunted by a 5 pF capacitor (not shown). The components within the dotted box were fabricated on-chip.

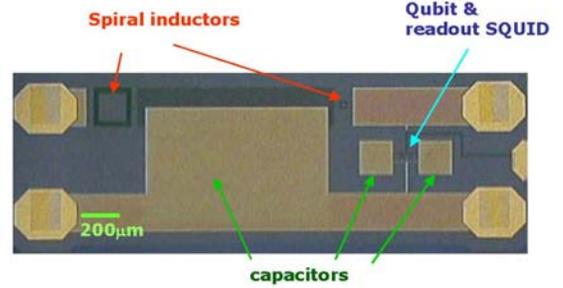

Fig. 4. Optical micrograph of the actual device fabricated at MIT Lincoln Laboratory.

## III. MEASUREMENT TECHNIQUES

The device was measured in a $^3$He refrigerator at MIT Lincoln Laboratory. The DC lines were made from shielded soft-coaxes and were filtered at the $^3$He pot sample stage with copper powder filters. The RF lines were attenuated at both the 1K and $^3$He stages. A bias-tee was mounted on the sample stage to combine the DC and RF signals. While the RF lines were used for the resonant readout, the DC lines were used to characterize the I-V properties of the SQUID junctions and to provide the option of additional DC bias to the SQUID. The chip was mounted on a PCB substrate with coplanar-waveguide structures and was housed inside an RF package. A superconducting coil was wrapped around the outside of the RF package to provide the external flux bias. An additional thermometer was mounted directly on the package to monitor the sample temperature.

A schematic of the experimental implementation is shown in Fig. 5. The frequency spectra of the readout circuit were measured with a network analyzer and/or a spectrum analyzer as a function of the input power from a tracking generator. The signal frequencies were centered about the resonant frequency and within a span of 0.6 MHz. Each spectrum was acquired with a RBW of 3 kHz and then averaged 100 times. The signal from the resonant circuit was amplified at room temperature with a 50 dB LNA before being measured.

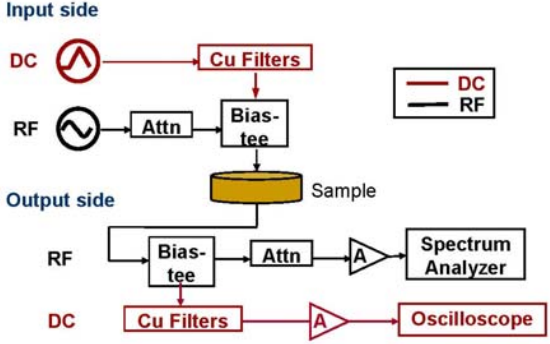

Fig. 5. Circuit schematic of the experimental setup.

## IV. EXPERIMENTAL RESULTS

### A. Readout of Qubit States

The resonant frequency of the readout circuit was measured to be about 420 MHz while the quality factor Q is on the order of a thousand. The observed Q is higher than the designed value; the discrepancy was likely due to the higher inductances of the spiral inductors. The resonant spectra were modulated with an external magnetic field. The periodic modulation of the resonant frequency and the corresponding peak amplitude are shown in Fig. 6. A discontinuity in the frequency is evidenced at every 1.3 periods of the SQUID lobe (lower trace, Fig. 6). These steps correspond to transitions of the qubit between opposite flux states, while the periodicity is determined by the ratio of the SQUID to qubit loop areas. The shift in the resonant frequency was about 40 kHz, corresponding to a change in Josephson inductance of 2 pH. While the data shown in Fig. 6 were obtained by measuring the magnitude of the resonant spectra, we have also detected the state of the qubit by measuring the phase of the resonant readout circuit with comparable sensitivity.

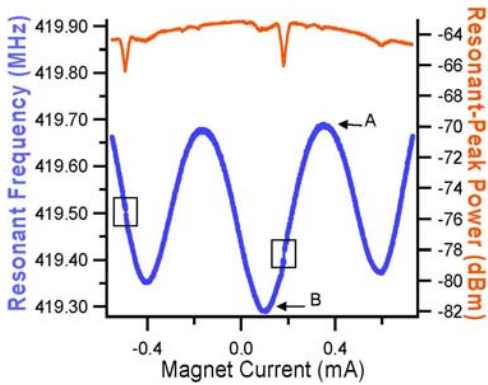

Fig. 6. The lower plot (left axis) shows the modulation of the resonant frequency with external magnetic field. Qubit steps corresponding to transitions between opposite flux states were observed at every 1.3 periods of the SQUID lobe. The upper plot (right axis) shows the corresponding peak amplitude of the resonant spectrum. The dip in peak power coincides with the qubit step.

A dip in resonant peak power was also observed to coincide with the qubit steps (upper trace, Fig. 6). This dip corresponds to a broadening of the resonant spectrum (not shown). To further understand the nature of the dip in the resonant peak power, we performed the measurements over a range of input power near one of the qubit steps (Fig. 7). For low current amplitudes, the resonant peak power remains fairly constant, and there is no dip at the qubit step. As one increases the current amplitude, the size of the dip grows and is eventually washed out at the highest biases. For all values of current amplitudes, the qubit step in the resonant frequency was still clearly observed. These preliminary results are consistent with the Landau-Zener transition data observed in Ref. [10] in the large bias limit. The temperature dependence of the results is to be further explored at dilution refrigerator temperatures.

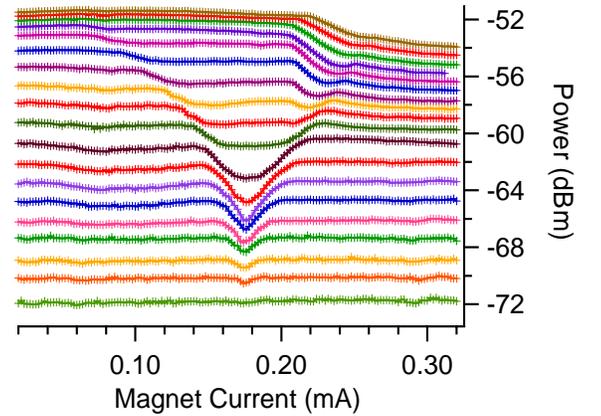

Fig. 7. Observation of the resonant peak power with increasing input power near a qubit step. The source power ranges from -68 dBm to -50 dBm in 1dB step. A dip starts to develop at the flux value corresponding to the position of the qubit step, and is gradually washed out with increasing power.

### B. Nonlinearity of Resonant Readout Circuit

As the bias current through the SQUID is increased along the supercurrent branch, the Josephson inductance becomes increasingly nonlinear and has observable effects on the resonant circuit spectra. Fig. 8 shows the behavior of the magnitude and phase spectra with increasing input power at the top of the SQUID lobe (position [A], Fig. 6). The magnitude spectrum evolves from a symmetric Lorenzian shape (linear oscillator) to an asymmetric shape with a discontinuity (nonlinear oscillator). The resonant spectrum leans towards lower frequencies, indicating that the effective inductance over an oscillation period is higher with increasing current bias. On the other hand, the phase experiences a 180° shift at the resonant frequency. At low current biases and thus in the linear regime, the phase shift is continuous with frequency and has a finite slope at resonance that is limited by the quality factor. In the nonlinear case, the phase spectra exhibit a discontinuity similar to the magnitude spectra. The discontinuity corresponds to a bifurcation point at which the nonlinear resonant circuit has two stable solutions. Since the data shown in Fig. 8 were only measured with a forward frequency sweep (i.e., from low to high frequencies), the spectra exhibit only one of the two stable branches.

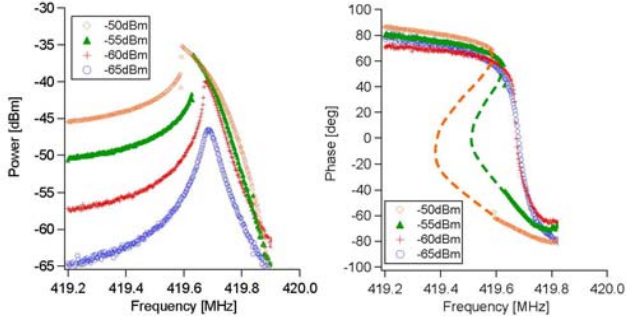

Fig. 8. Magnitude and phase measurements of the resonant readout circuit in the linear and nonlinear regimes at the top of the SQUID lobe. In the linear case, the magnitude spectrum has a symmetric Lorenzian shape, while the phase at resonance is continuous with a slope limited by the quality factor Q. In the nonlinear case, the magnitude and phase spectra have two stable branches, resulting in a discontinuity. The theoretical predictions are shown in dotted line for the phase spectra.

Fig. 9 shows the hysteretic magnitude spectra with both stable branches measured using forward and backward frequency sweeps. The third branch (displayed as a white dotted line) is unstable and inaccessible without the implementation of a feedback scheme. The behavior of the spectra at two different magnetic field biases is also shown in Fig. 9. The spectrum evolves from leaning towards the lower frequencies to leaning towards the higher frequencies as the flux bias changes from the top to the bottom of the SQUID lobe.

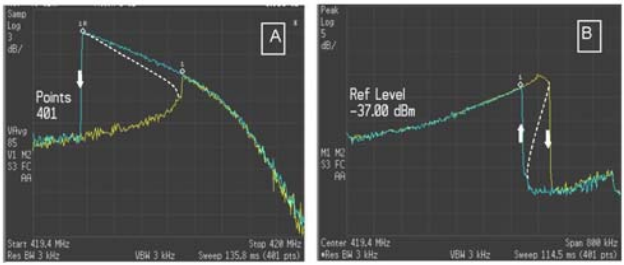

Fig. 9. Hysteretic behavior of the resonant spectrum was observed when the readout circuit was operated in the nonlinear regime. The results are shown for two magnetic flux biases: top [A] and bottom [B] of the SQUID lobe. The two stable branches were measured directly. The unstable branch was not measured, but is illustrated with a white dotted line.

The characterization of the readout scheme presented here involved measuring the magnitude or phase spectra and, subsequently, identifying the qubit transition as a shift in the resonant frequency. While this approach was beneficial for characterizing the resonant readout circuit, its implementation as a qubit readout scheme will likely involve pulsing the resonant circuit at a single frequency near the resonant frequency. Pulsing allows for a temporal resolution of the qubit state measurement, and the qubit state would be determined as a change in either the output amplitude or phase within the pulse. The circuit quality factor will determine the rate at which this information can be obtained. From this perspective, the readout circuit operated in the nonlinear regime has the potential to be used as a nonlinear amplifier. One can choose to bias near the discontinuity point and benefit from the high sensitivity [7].

## V. Conclusion

We have implemented a resonant readout scheme for the persistent-current qubit by using the readout SQUID as a flux-sensitive inductor. This approach only requires low current biases through the SQUID and reduces the level of decoherence of the qubit as induced by the readout process. The results presented were measured from an on-chip niobium device at 300mK, and the readout scheme was confirmed to have the sensitivity to distinguish the two qubit states. The readout circuit was also characterized in both the linear and nonlinear regime. Future experiments involving the spectroscopy of the qubit states using the resonant approach will be performed at dilution refrigerator temperatures.

## Acknowledgment

The authors would like to acknowledge D. E. Oates, K. Segall, D. S. Crankshaw, and Y. Yu for useful discussions. Thanks also go to T. J. Weir for technical assistance.